\RequirePackage{lineno} 
\documentclass[aps,prl,letterpaper,twocolumn, floatfix,superscriptaddress]{revtex4}%
\usepackage{graphicx}
\usepackage{bm}
\usepackage{epsf}
\usepackage{rotating}
\usepackage{epsfig,graphics,rotate,color}
\usepackage{wrapfig}
\usepackage{amssymb}
\usepackage{amsmath}
\usepackage{amsfonts}
\usepackage{array,hhline,dcolumn}%
\usepackage{gensymb}
\usepackage{placeins}
\usepackage{pdfpages}
\usepackage[colorlinks=true,citecolor=blue,linkcolor=blue]{hyperref}
\usepackage{accents}
\usepackage{color}
\usepackage{xcolor}
\usepackage[normalem]{ulem}

\usepackage[utf8]{inputenc}
\providecommand{\U}[1]{\protect\rule{.1in}{.1in}}
\def\al{\alpha}

\def\ga{\gamma}

\def\th{\theta}

\def\la{\lambda}

\def\ph{\phi}

\def\Ph{\Phi}

\def\nue{\nu_e}
\def\numu{\nu_\mu}
\def\nutau{\nu_\tau}

\def\numubar{\bar\nu_\mu}

\def\fr#1#2{\frac{#1}{#2}}

\def\ket#1{|{#1}\rangle}

\def\lsim{\mathrel{\rlap{\lower4pt\hbox{\hskip1pt$\sim$}}
    \raise1pt\hbox{$<$}}}
\def\gsim{\mathrel{\rlap{\lower4pt\hbox{\hskip1pt$\sim$}}
    \raise1pt\hbox{$>$}}}
\def\Re{\hbox{Re}\,}
\def\Im{\hbox{Im}\,}

\newcommand{\beq}{\begin{eqnarray}}
\newcommand{\eeq}{\end{eqnarray}}
\def\to{\rightarrow}

\def\no{\nonumber}

\def\ismeadt{\accentset{\circ}{a}^{(d)}_{\mu\mu}} 
\def\ismecdt{\accentset{\circ}{c}^{(d)}_{\mu\mu}}

\def\ismegt{\accentset{\circ}{c}^{(6)}_{\mu\mu}}

\def\ismead{\accentset{\circ}{a}^{(d)}_{\mu\tau}} 
\def\ismecd{\accentset{\circ}{c}^{(d)}_{\mu\tau}} 
\def\ismea{\accentset{\circ}{a}^{(3)}_{\mu\tau}} 
\def\ismec{\accentset{\circ}{c}^{(4)}_{\mu\tau}} 
\def\ismet{\accentset{\circ}{a}^{(5)}_{\mu\tau}}
\def\ismeg{\accentset{\circ}{c}^{(6)}_{\mu\tau}}
\def\ismes{\accentset{\circ}{a}^{(7)}_{\mu\tau}}
\def\ismej{\accentset{\circ}{c}^{(8)}_{\mu\tau}}

\def\ismeadn{\accentset{\circ}{a}^{(d)}} 
\def\ismecdn{\accentset{\circ}{c}^{(d)}} 
\def\ismean{\accentset{\circ}{a}^{(3)}} 
\def\ismecn{\accentset{\circ}{c}^{(4)}} 
\def\ismetn{\accentset{\circ}{a}^{(5)}}
\def\ismegn{\accentset{\circ}{c}^{(6)}}
\def\ismesn{\accentset{\circ}{a}^{(7)}}
\def\ismejn{\accentset{\circ}{c}^{(8)}}

\def\costhg{\ismegt/\rho_6}
\def\costha{\ismeadt/\rho_d}
\def\costhc{\ismecdt/\rho_d}

\def\NDOM{5160}
\def\NString{86}
\def\BString{1450}
\def\EString{2450}
\def\DHole{125}

\def\Nevt{34975}
\def\NbinE{17}
\def\lowElim{400}
\def\highElim{18}
\def\NbinUZ{10}
\def\lowUZlim{-1.0}
\def\highUZlim{0.0}

\def\nsys{\textrm{six}}

\def\cnorme{40}

\def\pikre{10}

\def\crindexe{2}
\def\aindexe{25}

\def\lima{2.0\times 10^{-24}}
\def\limc{2.7\times 10^{-28}}
\def\limt{1.5\times 10^{-32}}
\def\limg{9.1\times 10^{-37}}
\def\lims{3.6\times 10^{-41}}
\def\limj{1.4\times 10^{-45}}
\def\limann{2.9\times 10^{-24}}
\def\limcnn{3.9\times 10^{-28}}
\def\limtnn{2.3\times 10^{-32}}
\def\limgnn{1.5\times 10^{-36}}
\def\limsnn{8.3\times 10^{-41}}
\def\limjnn{5.2\times 10^{-45}}

\begin{document}

\title{Neutrino Interferometry for High-Precision Tests of Lorentz Symmetry with IceCube}

\author{M.~G.~Aartsen}
\affiliation{Department of Physics, University of Adelaide, Adelaide, 5005, Australia}
\author{M.~Ackermann}
\affiliation{DESY, D-15738 Zeuthen, Germany}
\author{J.~Adams}
\affiliation{Dept.~of Physics and Astronomy, University of Canterbury, Private Bag 4800, Christchurch, New Zealand}
\author{J.~A.~Aguilar}
\affiliation{Universit\'e Libre de Bruxelles, Science Faculty CP230, B-1050 Brussels, Belgium}
\author{M.~Ahlers}
\affiliation{Niels Bohr Institute, University of Copenhagen, DK-2100 Copenhagen, Denmark}
\author{M.~Ahrens}
\affiliation{Oskar Klein Centre and Dept.~of Physics, Stockholm University, SE-10691 Stockholm, Sweden}
\author{I.~Al~Samarai}
\affiliation{D\'epartement de physique nucl\'eaire et corpusculaire, Universit\'e de Gen\`eve, CH-1211 Gen\`eve, Switzerland}
\author{D.~Altmann}
\affiliation{Erlangen Centre for Astroparticle Physics, Friedrich-Alexander-Universit\"at Erlangen-N\"urnberg, D-91058 Erlangen, Germany}
\author{K.~Andeen}
\affiliation{Department of Physics, Marquette University, Milwaukee, WI, 53201, USA}
\author{T.~Anderson}
\affiliation{Dept.~of Physics, Pennsylvania State University, University Park, PA 16802, USA}
% ...


\collaboration{IceCube Collaboration}
\thanks{\begin{widetext}Corresponding authors email: \href{mailto:analysis@icecube.wisc.edu}{analysis@icecube.wisc.edu}.\end{widetext}}
\noaffiliation



\begin{abstract}
Lorentz symmetry is a fundamental space-time symmetry underlying both the Standard Model of particle physics and general relativity. This symmetry guarantees that physical phenomena are observed to be the same by all inertial observers. However, unified theories, such as string theory, allow for violation of this symmetry by inducing new space-time structure at the quantum gravity scale. Thus, the discovery of Lorentz symmetry violation could be the first hint of these theories in Nature. Here we report the results of the most precise test of space-time symmetry in the neutrino sector to date. We use high-energy atmospheric neutrinos observed at the IceCube Neutrino Observatory to search for anomalous neutrino oscillations as signals of Lorentz violation. We find no evidence for such phenomena. This allows us to constrain the size of the dimension-four operator in the Standard-Model Extension for Lorentz violation to the $10^{-28}$ level and to set limits on higher dimensional operators in this framework. These are among the most stringent limits on Lorentz violation set by any physical experiment. 

\end{abstract}

\maketitle

\section{Introduction}

Very small violations of Lorentz symmetry, or ``Lorentz violation'' (LV), are allowed in many ultra-high-energy theories, including string theory~\cite{Kostelecky:1988zi}, noncommutative field theory~\cite{Carroll:2001ws}, and supersymmetry~\cite{GrootNibbelink:2004za}. The discovery of LV could be the first indication of such new physics.  Because of this, there are world-wide efforts underway to search for evidence of LV. The Standard-Model Extension (SME) is an effective-field-theory framework to systematically study LV~\cite{Kostelecky:2011gq}. The SME includes all possible types of LV which respect other symmetries of the Standard Model such as energy-momentum conservation and coordinate independence. Thus, the SME can provide a framework to compare results of LV searches from many different fields such as photons~\cite{Nagel:2014aga,Komatsu:2008hk,Kostelecky:2013rv,Kostelecky:2016kkn}, nucleons~\cite{Maccione:2009ju,Allmendinger:2013eya,Smiciklas:2011xq}, charged leptons~\cite{Heckel:2006ww,Bennett:2007aa,Pruttivarasin:2014pja}, and gravity~\cite{Kostelecky:2015dpa}. Recently, neutrino experiments have performed searches for LV~\cite{Abbasi:2009nfa,Abbasi:2010kx,Abe:2014wla}. So far all searches have obtained null results. The full list of existing limits from all sectors and a brief overview of the field are available elsewhere~\cite{Kostelecky:2008ts,Liberati:2013xla}. Our focus here is to present the most precise test of LV in the neutrino sector. 

The fact that neutrinos have mass has been established by a series of experiments ~\cite{Fukuda:1998mi,Ahmad:2001an,Ahn:2002up,Abe:2011sj,Eguchi:2002dm,An:2012eh}. The field has incorporated these results into the ``neutrino Standard Model’’($\nu$SM)—the SM with three massive neutrinos.   Although the $\nu$SM parameters are not yet fully determined~\cite{Esteban:2016qun}, the model is rigorous enough to be brought to bear on the question of LV. We briefly review in Methods the history of neutrino oscillation physics and tests of Lorentz violation with neutrinos.

To date, neutrino masses have proved to be too small to be measured kinematically, but the mass differences are known via neutrino oscillations. This phenomenon arises from the fact that production and detection of neutrinos involves the flavour states, while the propagation is given by the Hamiltonian eigenstates. Thus, a neutrino with flavour $\ket{\nu_\al}$ can be written as a superposition of Hamiltonian eigenstates $\ket{\nu_i}$, i.e., $\ket{\nu_\alpha}= \sum_{i=1}^3 V_{\alpha i}(E)\ket{\nu_i}$, where $V$ is the unitary matrix that diagonalizes the Hamiltonian and in general is a function of neutrino energy $E$. When the neutrino travels in vacuum without new physics, the Hamiltonian depends only on the neutrino masses, and the Hamiltonian eigenstates coincide with the mass eigenstates. That is, $H=\frac{1}{2E}\cdot U^{\dagger}{\rm diag}(m^2_1,m^2_2,m^2_3)U=\frac{m^2}{2E}$, where $m_i$ are the neutrino masses and $U$ is the 
PMNS matrix which diagonalizes the mass matrix $m$~\cite{Esteban:2016qun}. 

A consequence of the flavour misalignment is that a neutrino beam that is produced purely of one flavour will evolve to produce other flavours.  Experiments measure the number of neutrinos of different flavours, observed as a function of the reconstructed energy of the neutrino, $E$, and the distance the beam has traveled, $L$. The microscopic neutrino masses are directly tied to the macroscopic neutrino oscillation length. In this sense, neutrino oscillations are similar to photon interference experiments in their ability to probe very small scales in Nature.

\section{Lorentz violating neutrino oscillations}

\begin{figure}
\centering
\includegraphics[width=1.0\columnwidth]{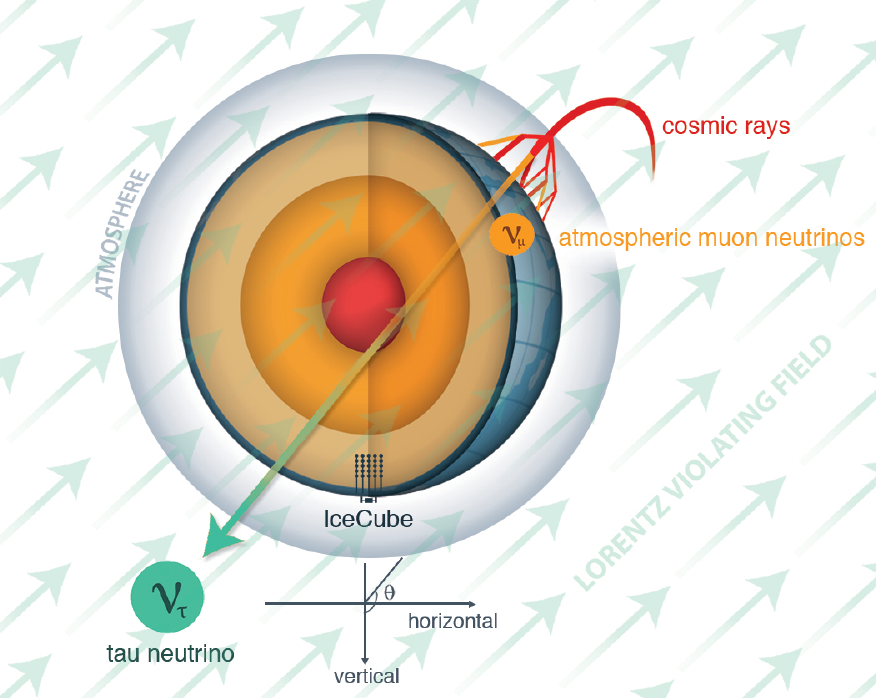}
\caption{Schematic figure of the test of LV with atmospheric neutrinos in IceCube. Muon neutrinos produced in the upper atmosphere are detected by IceCube in Antarctica. The potential signal is the anomalous disappearance of muon neutrinos, which might be caused by the presence of a hypothetical LV field that permeates space. The effect can be directional (arrows), but in this analysis we test the isotropic component.}
\label{fig:cartoon}
\end{figure}

In this analysis, we use neutrino oscillations as a natural interferometer with a size equal to the diameter of the Earth. We look for anomalous flavour-changing effects caused by LV that would modify the observed energy and zenith angle distributions of atmospheric muon neutrinos observed in the IceCube Neutrino Observatory~\cite{Aartsen:2015rwa} (see Figure~\ref{fig:cartoon}). Beyond flavour change due to small neutrino masses, any hypothetical LV fields could contribute to muon neutrino flavour conversion. Thus, in this analysis, we look for distortion of the expected muon neutrino distribution. Since this analysis does not distinguish between a muon neutrino ($\numu$) and its antineutrino ($\numubar$), when the word ``neutrino" is used, we are referring to both.

Past searches for LV have mainly focused on the directional effect in the Sun-centred celestial-equatorial frame (SCCEF)~\cite{Kostelecky:2008ts} by looking only at the time dependence of physics observables as direction-dependent physics appears as a function of the Earth's rotation.  However, in our case, we assume no time dependence, and instead look at the energy distribution distortions caused by direction- and time-independent isotropic LV. Isotropic LV may be a factor $\sim 10^{3}$ larger than direction-dependent LV in SCCEF if we assume that the new physics is isotropic in the CMB frame~\cite{Liberati:2013xla}. It would be most optimal to simultaneously look for both effects, but our limited statistics do not allow for this.

To calculate the effect, we start from an effective Hamiltonian derived from the SME~\cite{Kostelecky:2011gq}, which can be written as 
\beq
H\sim\frac{m^2}{2E}+\ismean-E\cdot\ismecn+E^2\cdot\ismetn-E^3\cdot\ismegn\cdots .
\label{eq:Heff}
\eeq
The first term of Eq.~\eqref{eq:Heff} is from the $\nu$SM, however, its impact decreases at high energy. The remaining terms ($\ismean$, $\ismecn$, $\ismetn$, and so on) arise from the SME and describe isotropic Lorentz violating effects. The circle symbol on top indicates  isotropic coefficients, and the number in the bracket is the dimension of the operator.  These terms are typically classified as CPT-odd ($\ismeadn$) and CPT-even ($\ismecdn$). Focusing on muon neutrino to tau neutrino ($\numu\rightarrow \nutau$) oscillations, all SME terms in Eq.~\eqref{eq:Heff} can be expressed as $2\times 2$ matrices, such as
\beq
\ismegn=
\left(\begin{array}{cc}
\ismegt & \ismeg \\ 
{\ismeg}^* & 
-\ismegt
\end{array}\right)~.
\label{eq:matrix}
\eeq
Without loss of generality, we can define the matrices so that they are traceless, leaving three independent parameters, in this case: $\ismegt$, $\Re(\ismeg)$, and $\Im(\ismeg)$. The off-diagonal Lorentz violating term $\ismeg$ dominates neutrino oscillations at high energy, which is the main interest of this paper. In this formalism, LV can be described by an infinite series, but higher order terms are expected to be suppressed. Therefore, most terrestrial experiments focus on searching for effects of dimension-three and -four operators; $\ismean$ and $E\cdot\ismecn$ respectively. However, our analysis extends to dimension-eight, {\it i.e.}, $E^2\cdot\ismetn$, $E^3\cdot\ismegn$, $E^4\cdot\ismesn$, and $E^5\cdot\ismejn$. Such higher orders are accessible by IceCube, which observes high-energy neutrinos where we expect an enhancement from the terms with dimension greater than four. In fact, some theories, such as noncommutative field theory~\cite{Carroll:2001ws} and supersymmetry~\cite{GrootNibbelink:2004za}, allow for LV to appear in higher order operators. As an example, we expect dimension-six new physics operators of order $\sim \frac{1}{M_P^2}\sim 10^{-38}$~GeV$^{-2}$ where $M_P$ is the Planck mass which is the natural energy scale of the unification of all matter and forces including gravity.  We assume that only one dimension is important at any given energy scale, because the strength of LV is expected to be different at different orders. 

We use the $\numu \rightarrow \nutau$ two-flavour oscillation scheme following Ref.~\cite{GonzalezGarcia:2005xw}. This is appropriate because we assume there is no significant interference with $\nue$. Details of the model used in this analysis are given in Methods. The oscillation probability is given by
\beq
P(\numu\rightarrow\nutau)=-4V_{\mu 1}V_{\mu 2}V_{\tau 1}V_{\tau 2}\sin^2\left(\cfrac{\la_2-\la_1}{2}L \right),
\label{eq:osc}
\eeq
where $V_{\alpha i}$ are the mixing matrix elements of the effective Hamiltonian (Eq.~\eqref{eq:Heff}), and $\la_i$ are its eigenvalues. Both mixing matrix elements and eigenvalues are a function of energy, $\nu$SM oscillation parameters, and SME coefficients. Full expressions are given in Supplementary material Appendix~A.

\section{The IceCube neutrino observatory} 

The IceCube Neutrino Observatory is located at the geographic South Pole~\cite{Abbasi:2008aa,Aartsen:2016nxy}. The detector volume is one cubic kilometer of clear Antarctic ice. Atmospheric muon neutrinos interacting on surrounding ice or bedrock may produce high-energy muons, which emit photons that are subsequently detected by digital optical modules (DOMs) embedded in the ice. The DOMs consist of a 25~cm diameter Hamamatsu photomultiplier tube, with readout electronics, contained within a 36.5~cm glass pressure housing. These are installed in holes in the ice with roughly $\DHole$~m separation. There are $\NString$ holes in the ice with a total of $\NDOM$ DOMs, which are distributed at depths of $\BString$~m to $\EString$~m below the surface, instrumenting one gigaton of ice. The full detector description can be found in Ref.~\cite{Aartsen:2016nxy}.

This detector observes Cherenkov light from muons produced in charged-current $\numu$ interactions. Photons detected by the DOMs allow for the reconstruction of the muon energy and direction, which is related to the energy of the primary $\numu$.   Because the muons are above critical energy, their energy can be determined by measuring the stochastic losses that produce Cherenkov light.  See Ref.~\cite{Aartsen:2015rwa} for details on the muon energy proxy used in this analysis. In the TeV energy range, these muons traverse distances of order kilometers, and have small scattering angle due to the large Lorentz boost, resulting in 0.75$^\circ$ resolution on the reconstructed direction at 1~TeV~\cite{WeaverThesis}. We use up-going muon data of TeV-scale energy from two years of detector operation~\cite{Aartsen:2015rwa} representing $\Nevt$~events with a 0.1\% atmospheric muon contamination.  

\section{Analysis setup}

\begin{figure}
\centering
\includegraphics[width=1.0\columnwidth]{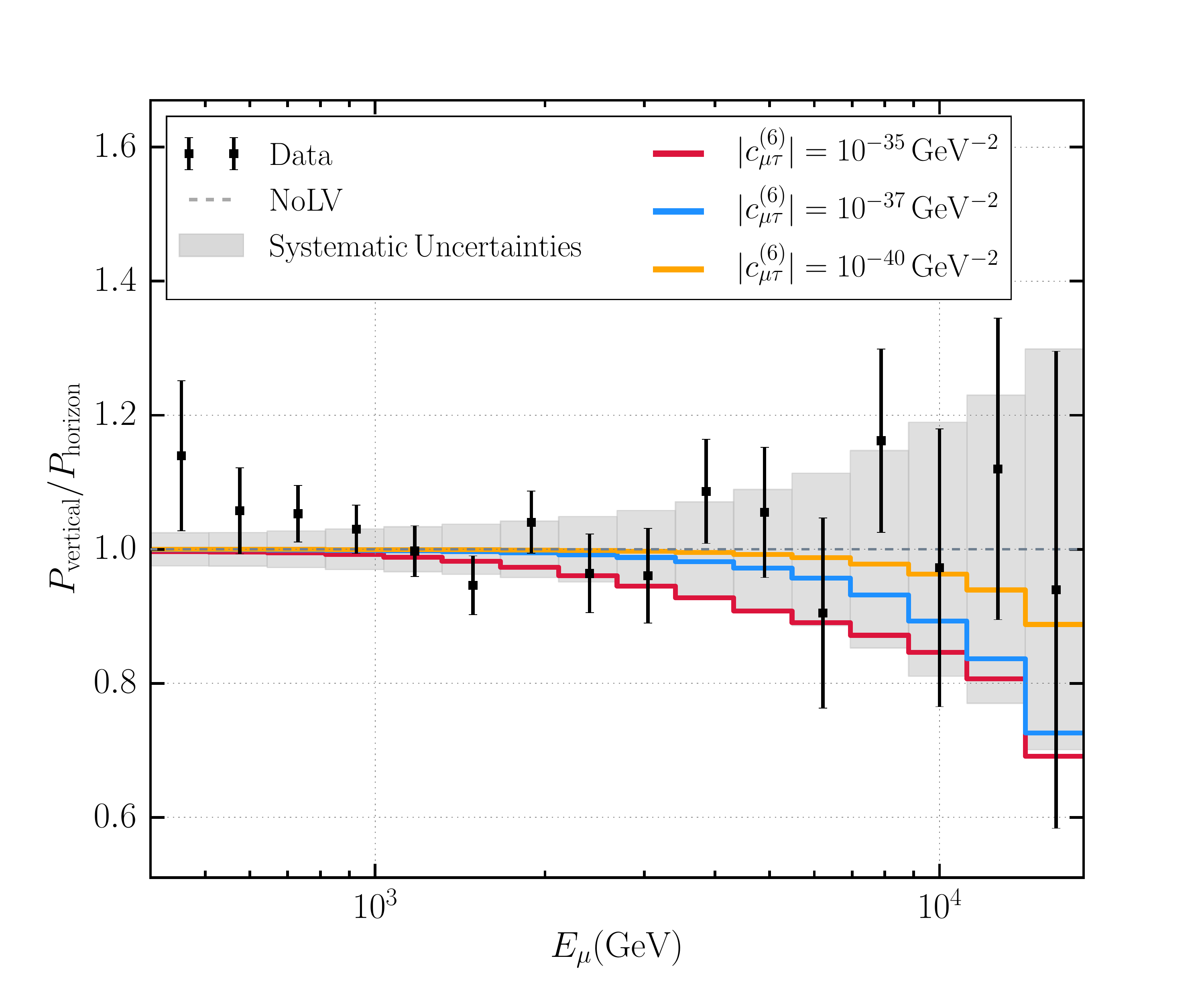}
\caption{The ratio of vertical to horizontal neutrino transition probabilities at IceCube. Here, vertical events are defined by $\cos\th\leq-0.6$ and the horizontal events are defined by $\cos\th>-0.6$. The transition probability ratio with one standard deviation statistical errors, are extracted from the data, is compared to the prediction for various dimension-six operator values: $10^{-35}$~GeV$^{-2}$ (red), $10^{-37}$~GeV$^{-2}$ (blue), and $10^{-40}$~GeV$^{-2}$ (yellow). The range of uncorrelated systematic uncertainties is shown as a light grey band. This is constructed from ensembles of many simulations where the nuisance  parameters are varied within their uncertainties.}
\label{fig:ratio}
\end{figure}

To obtain the prediction for LV effects, we multiply the oscillation probability, given in Eq.~\eqref{eq:osc}, with the predicted atmospheric neutrino flux calculated using the matrix cascade equation (MCEq)~\cite{Fedynitch:2015zma}. These ``atmospheric neutrinos'' originate from decays of muons and various mesons produced by collisions of primary cosmic rays and air molecules, and consist of both neutrinos and antineutrinos. The atmospheric neutrinos have two main components: ``conventional," from pion and kaon decay, and ``prompt," from charmed meson decay. The conventional flux dominates at energies less than $\highElim$~TeV because of the larger production cross section, whereas the harder prompt spectrum becomes relevant at higher energy. In the energy range of interest, the astrophysical neutrino contribution is small. We include it modelled as a power law with normalization and spectral index, $\sim\Ph\cdot E^{-\ga}$. The absorption of each flux component propagating through the Earth to IceCube is properly modelled~\cite{BenThesis,CarlosThesis}. Muon production from $\nu_\mu$ charged-current events at IceCube proceeds through deep inelastic neutrino interactions as calculated in Ref.~\cite{CooperSarkar:2007cv}.  

The short distance of travel for horizontal neutrinos leads to negligible spectral distortion due to LV, while the long pathlength for vertical neutrinos leads to modifications. Therefore, if we compare the zenith angle distribution ($\theta$) of the expectation from simulations and $\numu$ data from $\cos\theta=\lowUZlim$ (vertical) to $\cos\theta=\highUZlim$ (horizontal), see Fig.~\ref{fig:cartoon}, then one can determine the allowed LV parameters. Figure~\ref{fig:ratio} shows the ratio of transition probabilities of vertical events to horizontal events. The data transition probability is defined by the ratio of observed events to expected events, and the simulation transition probability is defined by the expected events in the presence of LV to the number of events in the absence of LV. In the absence of LV, this ratio equals one. Here, as an example, we show several predictions from simulations with different dimension-six LV parameters $|\ismeg |$. In general, higher order terms are more important at higher energies. In order to assess the existence of LV, we perform a binned Poisson likelihood analysis by binning the data in zenith angle and energy. We use $\NbinUZ$ linearly-spaced bins in cosine of zenith angle from $\lowUZlim$ to $\highUZlim$ and $\NbinE$ logarithmically-spaced bins in reconstructed muon energy ranging from $\lowElim$~GeV to $\highElim$~TeV. Systematic uncertainties are incorporated as nuisance parameters in our likelihood. We introduce $\nsys$ systematic parameters related to the neutrino flux prediction: normalizations of conventional ($\cnorme$\% error), prompt (no constraint), and astrophysical (no constraint) neutrino flux components; ratio of pion and kaon contributions for conventional flux ($\pikre$\% error); spectral index of primary cosmic rays ($\crindexe$\% error); and astrophysical neutrino spectral index ($\aindexe$\% error). The absolute photon detection efficiency has been shown to have negligible impact on the exclusion contours in a search for sterile neutrinos that uses an equivalent analysis technique for a subset of the IceCube data considered here~\cite{TheIceCube:2016oqi,BenThesis}. The impact of light propagation model uncertainties on the horizontal to vertical ratio is less than ~5\% at few TeV, where this analysis is most sensitive~\cite{CarlosThesis}. Thus the impact of these uncertainties on the exclusion contours are negligible.

To constrain the LV parameters we use two statistical techniques. First, we performed a likelihood analysis by profiling the likelihood over the nuisance parameters per set of LV parameters.  From the profiled likelihood, we find the best-fit LV parameters and derive the 90\% and 99\% confidence levels (C.L.) assuming Wilks' theorem with three degrees of freedom~\cite{TheIceCube:2016oqi}. Second, we set the priors to the nuisance parameter uncertainties and scan the posterior space of the likelihood by means of a Markov Chain Monte Carlo (MCMC) method~\cite{ForemanMackey:2012ig}. These two procedures are found to be complementary, and the extracted LV parameters agree with the null hypothesis. For simplicity, we present the likelihood results in this paper and show the MCMC results in Methods Supplementary material Appendix~B.

\section{Results}

\begin{figure}
\centering
\includegraphics[width=1.0\columnwidth]{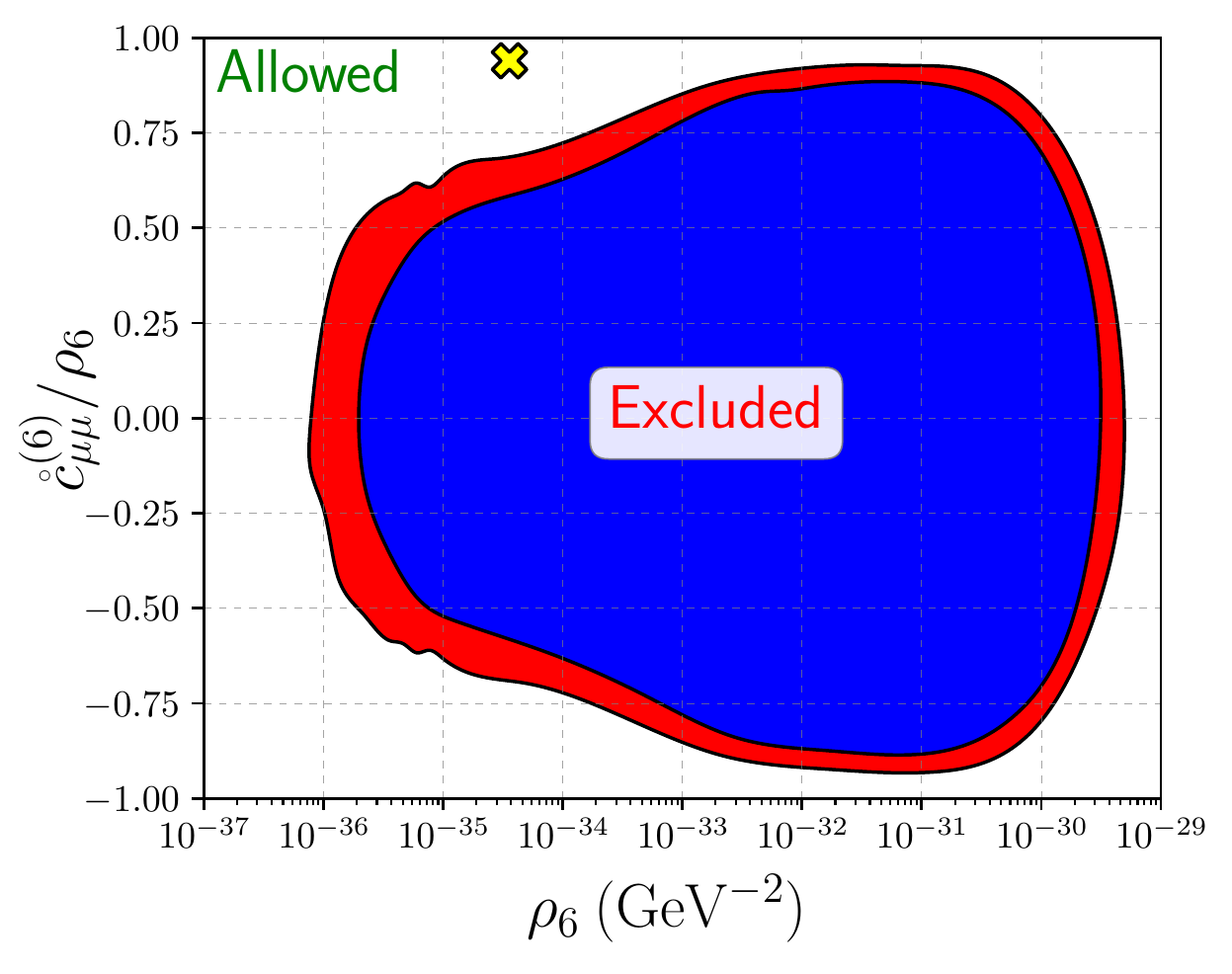}
\caption{The excluded parameter space region for the dimension-six SME coefficients. The parameters of horizontal and vertical axes are combinations of the three SME coefficients and explained in text. The best-fit point is shown by the yellow cross and the blue (red) region is excluded at 99\% (90\%) C.L.}
\label{fig:result_g_frequentist_limit}
\end{figure}

Figure~\ref{fig:result_g_frequentist_limit} shows the excluded region of dimension-six SME coefficients. The results for all operators are available in Supplementary material Appendix~C. 
The fit was performed in a three-dimensional phase space; however, the complex phase of the off-diagonal terms is not important at high energy, and we choose the following representation methods. The horizontal axis shows the strength of LV, $\rho_6\equiv\sqrt{(\ismegt)^2+\Re(\ismeg)^2+\Im(\ismeg)^2}$, and the vertical axis represents a fraction of the diagonal element, $\costhg$. The best-fit point shown by the marker is compatible with the absence of LV; therefore, we present 90\% C.L. (red) and 99\% C.L. (blue) exclusion regions. The contour extends to small values, beyond the phase space explored by previous analyses~\cite{Abbasi:2009nfa,Abbasi:2010kx,Abe:2014wla}. The leftmost edge of our exclusion region is limited by the small statistics of high-energy atmospheric neutrinos. The rightmost edge of the exclusion region is limited by fast LV-induced oscillations that suppress the flux but lead to no shape distortion. This can only be constrained by the absolute normalization of the flux. In the case of the dimension-three operator, the right edge can be excluded by other atmospheric neutrino oscillation measurements~\cite{Abe:2014wla,Aartsen:2014yll}. We have studied the applicability of Wilks' theorem via simulations. Near degenerate real and imaginary parameters reduce the expected degrees of freedom from three and the results here are interpreted as conservative confidence intervals.

Unlike previous results~\cite{Abbasi:2009nfa,Abbasi:2010kx,Abe:2014wla}, this analysis includes all parameter correlations, allowing for certain combinations of parameters to be unconstrained. This can be seen near $\costhg$=-1 and 1, where LV is dominated by the large diagonal component. This induces the quantum Zeno effect~\cite{Harris:1980zi}, where a neutrino flavour state is ``arrested" in one state by a continuous interaction with a LV field suppressing flavour transitions. Thus, the unshaded regions below and above our exclusion zone are very difficult to constrain with terrestrial experiments.

\begin{table*}[t]
    \centering
    \footnotesize
    \begin{tabular}{c c c c c c}
    dim. & method & type & sector & limits & ref.\\
    \hline \hline
    3& CMB polarization     &astrophysical& photon &$\sim 10^{-43}$ GeV &\cite{Komatsu:2008hk}  \\
     & He-Xe comagnetometer & tabletop    & neutron&$\sim 10^{-34}$ GeV&\cite{Allmendinger:2013eya} \\
     & torsion pendulum     & tabletop    &electron&$\sim 10^{-31}$ GeV&\cite{Heckel:2006ww}  \\
     & muon g-2             & accelerator & muon   &$\sim 10^{-24}$ GeV&\cite{Bennett:2007aa} \\
     & neutrino oscillation & atmospheric &neutrino&
     $|\Re(\ismea)|,|\Im(\ismea)|$
     \begin{tabular}{c}
     $<\limann$  GeV (99\% C.L.)\\
     $<\lima$ GeV (90\% C.L.)
     \end{tabular}&this work\\
    \hline
    4& GRB vacuum birefringence &astrophysical& photon & $\sim 10^{-38}$&\cite{Kostelecky:2013rv} \\
     & Laser interferometer     & LIGO        & photon & $\sim 10^{-22}$&\cite{Kostelecky:2016kkn} \\
     &Sapphire cavity oscillator& tabletop    & photon & $\sim 10^{-18}$&\cite{Nagel:2014aga} \\
     & Ne-Rb-K comagnetometer   & tabletop    & neutron& $\sim 10^{-29}$&\cite{Smiciklas:2011xq} \\
     & trapped Ca$^+$ ion~      & tabletop    &electron& $\sim 10^{-19}$&\cite{Pruttivarasin:2014pja} \\
     & neutrino oscillation     & atmospheric &neutrino&
     $|\Re(\ismec)|,|\Im(\ismec)|$
     \begin{tabular}{c}
     $<\limcnn$ (99\% C.L.)\\
     $<\limc$ (90\% C.L.)
     \end{tabular}&this work\\
    \hline
    5& GRB vacuum birefringence   &astrophysical& photon & $\sim 10^{-34}$ GeV$^{-1}$&\cite{Kostelecky:2013rv} \\
     &ultra-high-energy cosmic ray&astrophysical& proton & $\sim 10^{-22}$ to $10^{-18}$ GeV$^{-1}$&\cite{Maccione:2009ju} \\
     & neutrino oscillation       & atmospheric &neutrino&
     $|\Re(\ismet)|,|\Im(\ismet)|$
    \begin{tabular}{c}
     $<\limtnn$  GeV$^{-1}$ (99\% C.L.)\\
     $<\limt$ GeV$^{-1}$ (90\% C.L.)
     \end{tabular}&this work\\
    \hline
    6& GRB vacuum birefringene         &astrophysical& photon &$\sim 10^{-31}$ GeV$^{-2}$&\cite{Kostelecky:2013rv} \\
     & ultra-high-energy cosmic ray    &astrophysical& proton &$\sim 10^{-42}$ to $10^{-35}$ GeV$^{-2}$&\cite{Maccione:2009ju}  \\
     &gravitational Cherenkov radiation&astrophysical& gravity&$\sim 10^{-31}$ GeV$^{-2}$&\cite{Kostelecky:2015dpa} \\
     & neutrino oscillation            & atmospheric &neutrino&
     $|\Re(\ismeg)|,|\Im(\ismeg)|$
    \begin{tabular}{c}
     $<\limgnn$ GeV$^{-2}$ (99\% C.L.)\\
     $<\limg$ GeV$^{-2}$ (90\% C.L.)
     \end{tabular}&this work\\
    \hline
    7& GRB vacuum birefringence&astrophysical& photon &$\sim 10^{-28}$ GeV$^{-3}$&\cite{Kostelecky:2013rv} \\
     & neutrino oscillation    & atmospheric &neutrino&
     $|\Re(\ismes)|,|\Im(\ismes)|$
    \begin{tabular}{c}
     $<\limsnn$ GeV$^{-3}$ (99\% C.L.)\\
     $<\lims$ GeV$^{-3}$ (90\% C.L.)
     \end{tabular}&this work\\
    \hline
    8&gravitational Cherenkov radiation&astrophysical& gravity&$\sim 10^{-46}$ GeV$^{-4}$&\cite{Kostelecky:2015dpa} \\
     & neutrino oscillation            & atmospheric &neutrino&
     $|\Re(\ismej)|,|\Im(\ismej)|$
    \begin{tabular}{c}
     $<\limjnn$ GeV$^{-4}$ (99\% C.L.)\\
     $<\limj$ GeV$^{-4}$ (90\% C.L.)
     \end{tabular}&this work\\
    \hline \hline
    \end{tabular}
    \caption{Comparison of attainable best limits of SME coefficients in various fields.}
    \label{tab:result}
\end{table*}

Table~\ref{tab:result} summarizes the results of this work along with representative best limits. A comprehensive list of LV tests is available in~\cite{Kostelecky:2008ts}. To date, there is no experimental indication of LV, and all these experiments have maximized their limits by assuming that all but one of the SME parameters are zero~\cite{Kostelecky:2008ts}. Therefore, to make our results comparable with previous limits, we adopt the same convention. For this, we set the diagonal SME parameters to zero and focus on setting limits on the off-diagonal elements. The details of the procedure used to set limits are given in Supplementary material Appendix~D. 

Let us consider the limits from the lowest to highest order. Dimension-three and -four operators are included in the renormalizable sector of SME. These are the main focus of experiments using photons~\cite{Nagel:2014aga,Kostelecky:2016kkn}, nucleons~\cite{Allmendinger:2013eya,Smiciklas:2011xq}, and charged leptons~\cite{Bennett:2007aa,Heckel:2006ww,Pruttivarasin:2014pja}. Going beyond terrestrial experiments, limits arising from astrophysical observations provide strong constraints~\cite{Komatsu:2008hk,Kostelecky:2013rv}. Among the variety of limits coming from the neutrino sector, the attainable best limits are dominated by atmospheric neutrino oscillation analyses~\cite{Abbasi:2009nfa,Abbasi:2010kx,Abe:2014wla}, where the longest propagation length and the highest energies enable us to use neutrino oscillations as the biggest interferometer on the Earth. The results from our analysis surpass past ones due to the higher statistics of high-energy atmospheric neutrinos and our improved control of systematic uncertainties.   Using a traditional metric, which assumes neutrinos to be massless, we can recast our result as an upper limit on any deviation of the speed of massless neutrinos from the speed of light due to LV. That is less than $10^{-28}$ at 99\% CL. This is about an order of magnitude improvement over past analyses~\cite{Abbasi:2009nfa,Abbasi:2010kx,Abe:2014wla}, and is of the same order as the deviation in speed that is expected due to the known neutrino mass at the energies relevant for this analysis.

Searches of dimension-five and higher LV operators are dominated by astrophysical observations~\cite{Kostelecky:2013rv,Kostelecky:2015dpa,Maccione:2009ju}. Among them, ultra-high-energy cosmic rays (UHECRs) have the highest measured energy~\cite{Abraham:2008ru} and are used to set the strongest limits on dimension-six and higher operators~\cite{Maccione:2009ju}. However, these limits are sensitive to the composition of UHECRs, which is currently uncertain~\cite{Liberati:2013xla,Aab:2016htd}. These limits assume that the cosmic rays at the highest energies are protons, but if they are in fact iron nuclei, then the UHECR limits are significantly reduced. Our analysis sets the most stringent limits in an unambiguous way across all fields for the dimension-six operator. Such high-dimension operators are generic signatures of new physics~\cite{Weinberg}. For example, dimension-five operator is an attractive possibility to produce neutrino masses, and dimension-six operators represent new physics interactions which can, for example, mediate proton decay. Although LV dimension-six operators, such as $\ismeg$,  are well motivated by certain theories including noncommutative field theory~\cite{Carroll:2001ws} and supersymmetry~\cite{GrootNibbelink:2004za}, they have so far not been probed with elementary particles due to the lack of available high-energy sources. Thus, our work pushes boundaries on new physics beyond the Standard Model and general relativity.

\section{Conclusion}

We have presented a test of Lorentz violation with high-energy atmospheric muon neutrinos from IceCube. Correlations of the SME coefficients are fully taken into account, and systematic errors are controlled by the fit. Although we did not find evidence for LV, this analysis provides the best attainable limits on SME coefficients in the neutrino sector along with limits on the higher order operators.  Comparison with limits from other sectors reveals that this work provides  among the best attainable limits on dimension-six coefficients across all fields: from tabletop experiments to cosmology. This is a remarkable point that demonstrates how powerful neutrino interferometry can be in the study of fundamental space-time properties. 

Further improvements on the search for LV in the neutrino sector using IceCube will be possible when the astrophysical neutrino sample is included~\cite{Aartsen:2013jdh}. Such analyses~\cite{Stecker:2014oxa,Arguelles:2015dca} will require a substantial improvement of detector and flux systematic uncertainty evaluations~\cite{Aartsen:2013rt,Aartsen:2015xup}. 
In the near future, water-based neutrino telescopes such as KM3NeT~\cite{Adrian-Martinez:2016fdl} and the ten-times-larger IceCube-Gen2~\cite{Aartsen:2014njl} will be in a position to observe more astrophysical neutrinos. With the higher statistics and improved sensitivity, these experiments will have an enhanced potential for discovery of Lorentz violation. 


\subsection*{Data availability}

The data that were used in this study are available in the IceCube Public Data Access ``Astrophysical muon neutrino flux in the northern sky with 2 years of IceCube data~\cite{Aartsen:2015rwa}" at \url{http://icecube.wisc.edu/science/data/}.


\begin{acknowledgments}

We acknowledge the support from the following agencies:
USA – U.S. National Science Foundation-Office of Polar Programs,
U.S. National Science Foundation-Physics Division,
Wisconsin Alumni Research Foundation,
Center for High Throughput Computing (CHTC) at the University of Wisconsin-Madison,
Open Science Grid (OSG),
Extreme Science and Engineering Discovery Environment (XSEDE),
U.S. Department of Energy-National Energy Research Scientific Computing Center,
Particle astrophysics research computing center at the University of Maryland,
Institute for Cyber-Enabled Research at Michigan State University,
and Astroparticle physics computational facility at Marquette University;
Belgium – Funds for Scientific Research (FRS-FNRS and FWO),
FWO Odysseus and Big Science programmes,
and Belgian Federal Science Policy Office (Belspo);
Germany – Bundesministerium für Bildung und Forschung (BMBF),
Deutsche Forschungsgemeinschaft (DFG),
Helmholtz Alliance for Astroparticle Physics (HAP),
Initiative and Networking Fund of the Helmholtz Association,
Deutsches Elektronen Synchrotron (DESY),
and High Performance Computing cluster of the RWTH Aachen;
Sweden – Swedish Research Council,
Swedish Polar Research Secretariat,
Swedish National Infrastructure for Computing (SNIC),
and Knut and Alice Wallenberg Foundation;
Australia – Australian Research Council;
Canada – Natural Sciences and Engineering Research Council of Canada,
Calcul Québec,
Compute Ontario,
Canada Foundation for Innovation,
WestGrid,
and Compute Canada;
Denmark – Villum Fonden,
Danish National Research Foundation (DNRF);
New Zealand – Marsden Fund;
Japan – Japan Society for Promotion of Science (JSPS),
and Institute for Global Prominent Research (IGPR) of Chiba University;
Korea – National Research Foundation of Korea (NRF);
Switzerland – Swiss National Science Foundation (SNSF); 
United Kingdom - Science and Technology Facilities Council (STFC),
and The Royal Society.

\end{acknowledgments}

\section*{Author contributions}

The IceCube Collaboration designed, constructed and now operates the IceCube Neutrino Observatory. Data processing and calibration, Monte Carlo simulations of the detector and of theoretical models, and data analyses were performed by a large number of collaboration members, who also discussed and approved the scientific results presented here. The main authors of this manuscript were C. Arg\"{u}elles, A. Kheirakdish, G. Collin, S. Mandalia, J. Conrad, and T. Katori. It was reviewed by the entire collaboration before publication, and all authors approved the final version of the manuscript.

\section*{Competing financial interests}

The authors declare no competing financial interests.

\appendix


\ifx \standalonesupplemental\undefined
\setcounter{page}{1}
\setcounter{figure}{0}
\setcounter{table}{0}
\fi

\renewcommand{\thepage}{Supplementary Information -- S\arabic{page}}
\renewcommand{\figurename}{SUPPL. FIG.}
\renewcommand{\tablename}{SUPPL. TABLE}

\section*{Supplemental materials}

\subsection*{Neutrino oscillations and tests of Lorentz violation}

The field of neutrino oscillations has been developed through a series of measurements of Solar~\cite{
Altmann:2000ft,Abdurashitov:2002nt,Hosaka:2005um,Aharmim:2005gt,Arpesella:2008mt}, atmospheric~\cite{
Ashie:2004mr,Adamson:2014vgd,Aartsen:2017nmd}, reactor~\cite{Abe:2008aa,Abe:2011fz,Ahn:2012nd,An:2013zwz}, and accelerator neutrinos~\cite{Adamson:2014vgd,
Abe:2017uxa,Adamson:2017gxd}. In the early days, the cause of neutrino oscillations was not precisely known, and Lorentz violation was suggested as a possible source of neutrino flavour anomalies~\cite{Coleman:1998ti} and so tests of Lorentz violation with high-energy astrophysical sources started to generate a lot of interest~\cite{AmelinoCamelia:1997gz}. Subsequetly, the $L/E$ dependence of standard neutrino oscillations was measured~\cite{Ashie:2004mr}. Because the neutrino mass term in the effective Hamiltonian has a $1/E$ energy dependence, it was a strong indication that a nonzero neutrino mass is in fact the cause of neutrino oscillations, not Lorentz violation. 
Then, the focus of the community shifted to consider Lorentz violation to be a second order effect in neutrino oscillations, and so neutrino oscillation data has been used to look for small deviations due to Lorentz violation from the standard neutrino mass oscillations. 

One approach to look for LV is to use a model-independent effective field theory, such as the Standard-Model Extension (SME)~\cite{Colladay:1996iz,Colladay:1998fq}.
SME is widely accepted in communities from low-energy table top experiments to high-energy particle physics and cosmology, to search for Lorentz violation. This formalism incorporates various fundamental features of quantum field theories, such as energy-momentum conservation, observer Lorentz transformations, and spin-statistics, however it includes violations of particle Lorentz transformations. A number of neutrino oscillation data sets have been analyzed using this formalism, including LSND~\cite{Auerbach:2005tq}, MiniBooNE~\cite{AguilarArevalo:2011yi}, MINOS~\cite{Adamson:2008aa,Adamson:2010rn,Adamson:2012hp,Rebel:2013vc}, Double Chooz~\cite{Abe:2012gw,Diaz:2013iba}, 
SNO~\cite{Diaz:2016fqd}, T2K~\cite{Abe:2017eot}, as well as the aforementioned IceCube-40 and Super-Kamiokande. These experiments can be classified into two groups. First, the presence of a direction-dependent field induces direction-dependent physics. In particular, neutrino beam lines are fixed and so such direction-dependent physics would show up as a time-dependence of neutrino oscillation data~\cite{Auerbach:2005tq,AguilarArevalo:2011yi,Adamson:2008aa,Adamson:2010rn,Adamson:2012hp,Rebel:2013vc,Abe:2012gw,Diaz:2016fqd,Abe:2017eot}. Second, a search of Lorentz violation is possible even without assuming the presence of a spatial component, {\it i.e.} no time-dependent physics, by utilizing distortions of the spectrum~\cite{Diaz:2013iba}. The results presented in this article are based on this second approach.

\subsection*{Neutrino oscillation formula}

Here, we illustrate how to calculate the oscillation probability for the case with nonzero isotropic Lorentz violations, such as $\ismeadn$ and $\ismecdn$. The effective Hamiltonian relevant for oscillation is given by
\beq
H&\sim&\frac{m^2}{2E}+\sum_{d\geq 3} E^{d-3}(\ismeadn-\ismecdn)~\no.
\eeq
Note that $\ismeadn$ are nonzero for $d=odd$, and $\ismecdn$ are nonzero for $d=even$. We assume that either one of them is nonzero. We use $\numu\to\nutau$ two-flavour approximation which allows us to solve the time-dependent Schr\"{o}dinger equation analytically to derive the neutrino oscillation formula with neutrino masses and LV. This choice is allowed because a large matter potential "arrests" $\nue$ (quantum Zeno effect~\cite{Harris:1980zi}) and prevent transitions from $\numu$. Since the matter potential of $\nue$ is much bigger than that due to Lorentz violation effects, the size of Lorentz violation that we consider here hardly induces any $\numu\to\nue$ transition. Our choice of the two-flavour oscillation model does not diminish the strength of our constraints on parameters in the $\numu-\nutau$ block matrix with respect to a full three flavour calculation. Hence, the mass matrix $m^2$ can be diagonalized to $M^2=diag(m_2^2,m_3^2)$ by a mixing matrix $U$ with mixing angle $\ph$,
\beq
m^2
&=&UM^2U^{\dagger}\no\\ 
&=&\left(\begin{array}{cc}
\cos\ph & \sin\ph \\ 
-\sin\ph & \cos\ph
\end{array}\right)
\left(\begin{array}{cc}
m_2^2 & 0 \\ 
0 & m_3^2
\end{array}\right)
\left(\begin{array}{cc}
\cos\ph & -\sin\ph \\ 
\sin\ph & \cos\ph
\end{array}\right). \no
\eeq

By adding $E^{d-3}(\ismeadn-\ismecdn)$, 
this $2\times 2$ Hamiltonian can be diagonalized with two eigenvalues, $\la_1$ and $\la_2$. Here, we define $\la_2>\la_1$.
Then the oscillation formula is
\beq
P(\numu\to\nutau)
&=&\fr{|2A_2|^2}{(\la_2-\la_1)^2}\sin^2\left(\fr{\la_2-\la_1}{2}L\right)\no
\eeq
where
\beq
\la_{1}&=&\frac{1}{2}\left[
(A_1+A_3)-\sqrt{\left|A_1-A_3\right|^2+|2A_2|^2}\right] \no\\
\la_{2}&=&\frac{1}{2}\left[
(A_1+A_3)+\sqrt{\left|A_1-A_3\right|^2+|2A_2|^2}\right] \no\\
A_1&=&\frac{1}{2E}(m_2^2\cos^2\ph+m_3^2\sin^2\ph)+E^{d-3}(\ismeadt-\ismecdt) \no\\
A_2&=&\frac{1}{2E}\cos\ph \sin\ph(m_2^2-m_3^2)+E^{d-3}(\ismead-\ismecd) \no\\
A_3&=&\frac{1}{2E}(m_2^2\sin^2\ph+m_3^2\cos^2\ph)-E^{d-3}(\ismeadt-\ismecdt). \no
\eeq

In the high-energy limit, the neutrino mass effect is negligible in comparison to Lorentz violating effects, 
\beq
P(\numu\to\nutau)
&\sim&\left(
1-\fr{[\ismeadt-\ismecdt]^2}{\rho_d^2}
\right)
\sin^2(L\rho_d\cdot E^{d-3})\no\\
&=&
\fr{\left|\ismead-\ismecd\right|^2}{\rho_d^2}\sin^2(L\rho_d\cdot E^{d-3}).\no
\eeq
Here we use $\rho_d\equiv\sqrt{(\ismeadt)^2+\Re(\ismead)^2+\Im(\ismead)^2}$ or $\sqrt{(\ismecdt)^2+\Re(\ismecd)^2+\Im(\ismecd)^2}$, which represents the strength of LV. Then, $\ismeadt/\rho_d$ and $\ismecdt/\rho_d$ become fractions of diagonal terms which are bounded between $-1$ and $+1$. The result suggests there are no LV neutrino oscillations without off-diagonal terms and that the LV oscillations are symmetric between the real and imaginary parts of the off-diagonal SME parameters. 

\section*{Results using Bayesian framework\label{sec:mcmc}}

\begin{figure}[t]
\centering
\includegraphics[width=1.0\columnwidth]{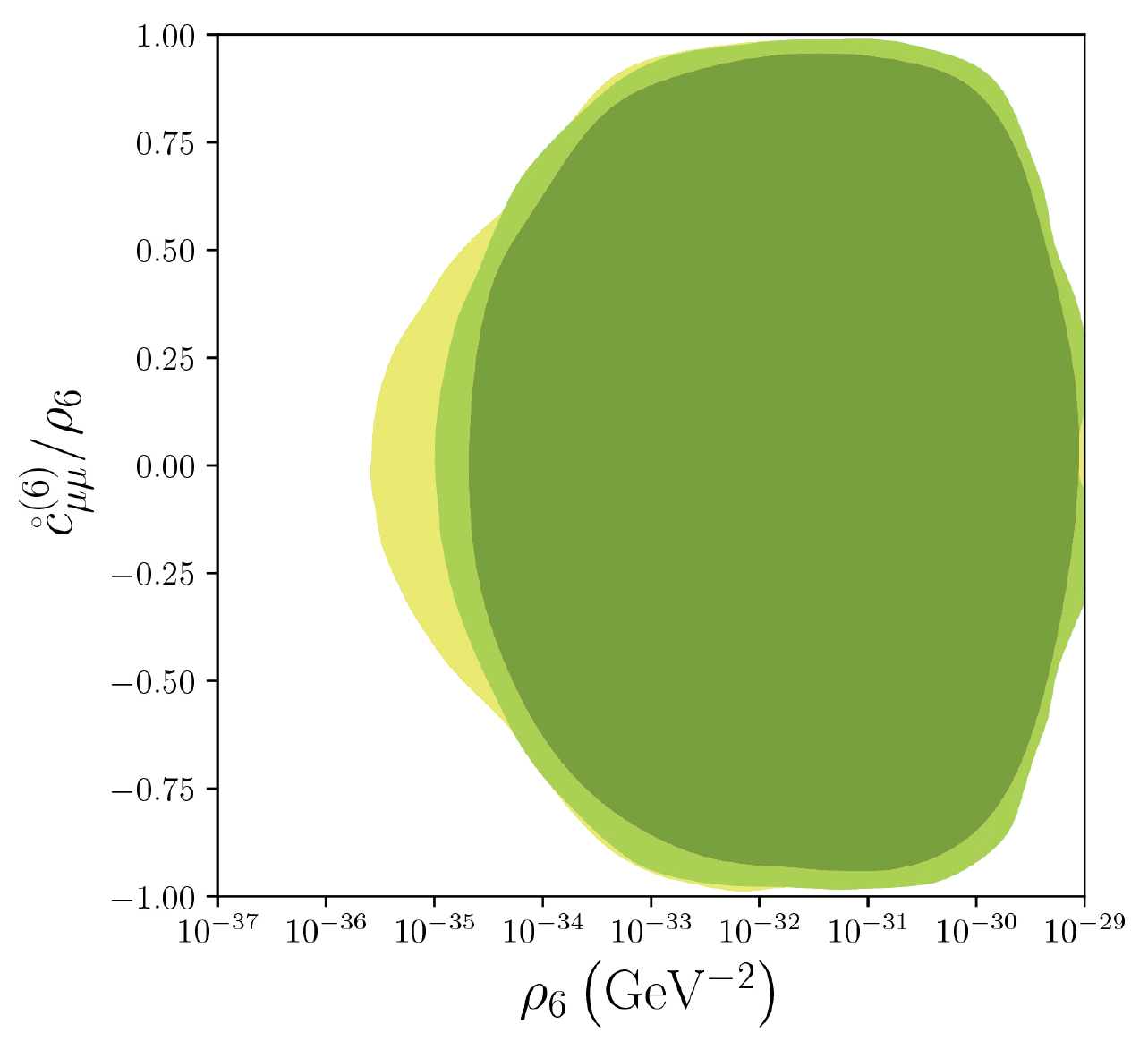}
\caption{The figure represents the posterior distribution on the dimension-six operator parameters marginalized (light green) or profiled (yellow) over all nuisance parameters. Both contours correspond to the exclusion with 99\% credibility region (C.R.). The dark green area corresponds to a Bayes factor $\mathcal{B} > 10^{1.5}$ with respect to the null hypothesis.}
\label{fig:credible_regions}
\end{figure}

The main results of this paper are extracted using Wilks' theorem so as to be directly comparable with frequentist results reported by other neutrino experiments. For completeness, we have also performed a Bayesian analysis which uses a joint distribution over the nine systematic and LV parameters. This joint distribution is constructed from the same likelihood and prior distributions used in the frequentist analysis, except that we also added conservative constraints on all flux normalizations to avoid a strong prior range dependence. The Bayesian study is presented in two results (see in Supplementary Figure~\ref{fig:credible_regions}), which were both generated by the EMCEE Markov Chain Monte Carlo software package~\cite{ForemanMackey:2012ig}. First, we constructed the 99\% exclusion credibility region (C.R.) from a sampling of the joint distribution, with two different treatments on nuisance parameters. Second, we extracted the result based on the Bayes factor of marginalizing the likelihood over nuisance parameters using the MultiNest algorithm~\cite{Feroz:2008xx}. These studies highlight the differences in results obtained using different treatments of nuisance parameters.

Supplementary Figure~\ref{fig:credible_regions} shows the example of Bayesian studies for dimension-six operators. First, we have constructed the 99\% exclusion credibility region (C.R.) from a sampling of the joint distribution, which is shown (light green contour). The shape differences observed between our main frequentist result and this one arise from the different statistical treatment of nuisance parameters. In a Bayesian analysis, nuisance parameters are marginalized instead of the profiling method used in a frequentist analysis. To demonstrate this, in Suppl. Fig.~\ref{fig:credible_regions}, we also show the credibility region where nuisance parameters are profiled (yellow contour). 

Second, we also report the Bayes factor ($\mathcal{B}$) as a function of the LV parameter space. The Bayes factor is the ratio of Bayesian evidence between the LV hypotheses and the no-LV hypothesis. In this result, the Bayesian evidence is found by marginalizing the likelihood over all parameters except for $\rho_d$ and $\costhg$. This is also shown in Supplementary Figure~\ref{fig:credible_regions}. According to Jeffreys' scale, a Bayes factor $\mathcal{B} > 10^{1.5}$ (dark green contour) is rejected with very strong strength-of-evidence. As expected, the Bayesian evidence based test has less reach with respect to the maximum likelihood ratio test used in the frequentist result, and it is consistent with the credibility region with marginalized nuisance parameters (light green contour). 

\section*{Full fit results from Wilks' theorem\label{sec:full}}

\begin{figure}[t]
\centering
\includegraphics[width=1.0\columnwidth]{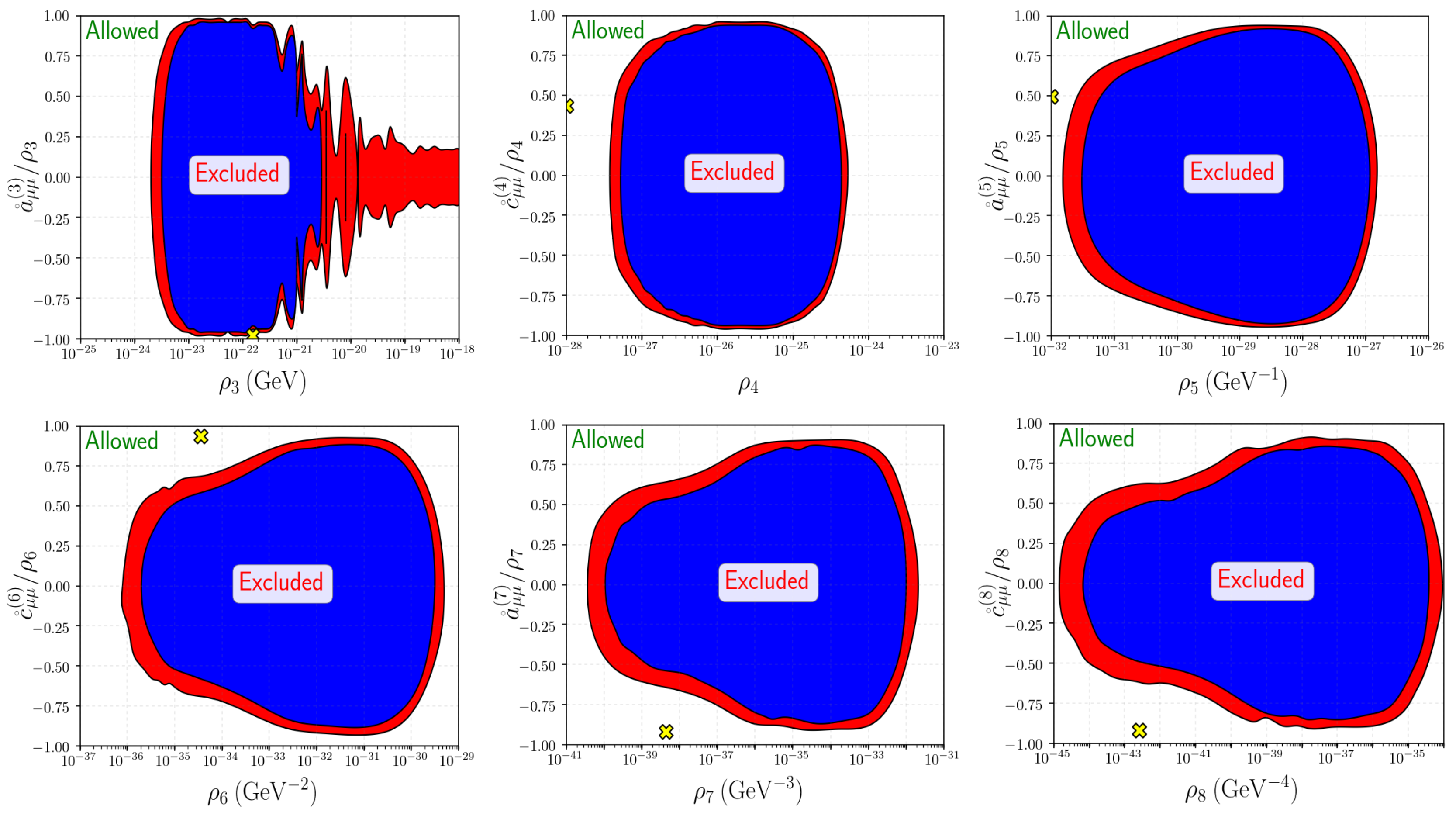}
\caption{These plots show the excluded parameter space with full parameter correlations. The x-axis represents the strength of the LV, and the y-axis shows the particular combination of SME coefficients. The dimension of the operator $d$ increases from 3 to 8 in these plots, from left to right, and top to bottom. The red (blue) regions are excluded at 90\% (99\%) C.L. as we discussed, near $cos\th_d=-1$ and $+1$, and at large values of $\rho^{(d)}$. The best-fit points are shown by the yellow crosses.}
\label{fig:result_fulllimit}
\end{figure}

Supplementary Figure~\ref{fig:result_fulllimit} shows the full-fit results from a two-flavour $\mu-\tau$ oscillation hypothesis with dimension-three to -eight LV operators. The x-axis represents the strength of LV, $\rho_d\equiv\sqrt{(\ismeadt)^2+\Re(\ismead)^2+\Im(\ismead)^2}$ or $\sqrt{(\ismecdt)^2+\Re(\ismecd)^2+\Im(\ismecd)^2}$, and the y-axis represents a fraction of the diagonal element, $\costha$ or $\costhc$.
The best-fit values indicate no LV, and we draw exclusion curves for 90\% C.L. (red) and 99\% C.L. (blue). The best fit points are shown with markers. If the best-fit points are outside the plotting region, markers are set at the edge. The overlap of contours from different dimension operators indicate that it is possible that two operators from different dimensions show up in the same energy scale.

Low dimension operators may be accessible by existing atmospheric neutrino oscillation experiments. For example, the DeepCore oscillation analysis utilizes $10-100$ GeV atmospheric neutrinos with a baseline equal to the Earth's diameter. This implies that the experiment has sensitivity to LV both via spectrum distortions and normalization changes of the flux, up to roughly a $10^{-23}$ GeV limit on $\ismean$ or $\rho_3$.

\begin{figure}[t]
\centering
\includegraphics[width=1.0\columnwidth]{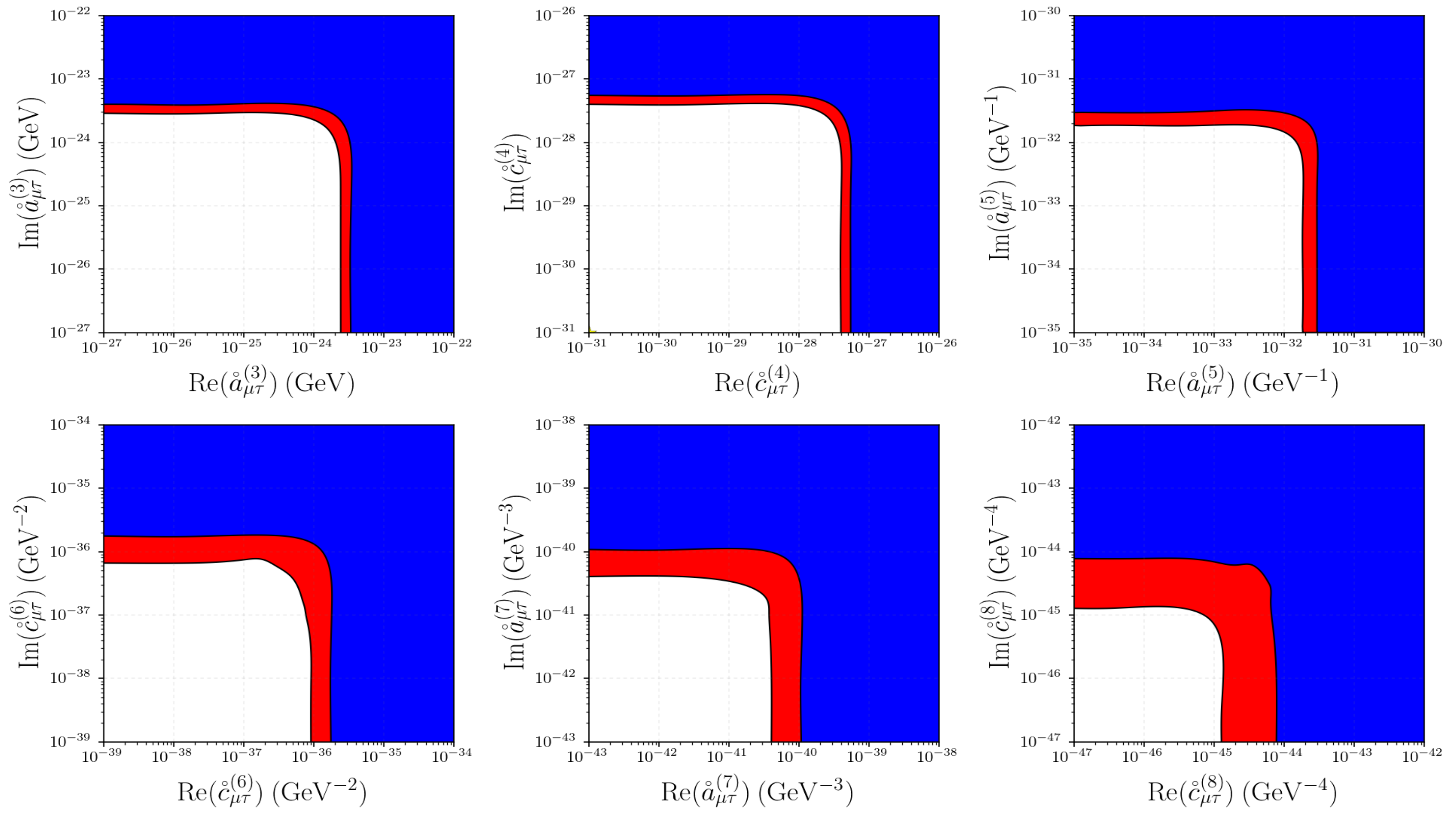}
\caption{These plots show limits on off-diagonal parameters in the case when diagonal parameters are set to zero. The dimension of the operator $d$ increases from 3 to 8 in these plots, from left to right, and top to bottom. The red (blue) regions are excluded at 90\% (99\%) C.L. There are four identical plots depending on the sign of the real and imaginary parts, but here we only show the cases when both the real and imaginary parts are positive.
}
\label{fig:result_bestlimit}
\end{figure}

\end{document}